\documentclass[12p]{article}
\usepackage{amssymb,amsmath,amsthm,latexsym,amsfonts,amscd,dsfont,enumerate,epsf,color}
\usepackage{a4wide,graphicx,tikz}

\usepackage[english]{babel}

\usepackage{color}

\newcommand{\sbar}{\nu}

\newcommand{\sigb}{\bar{\sigma}}

\newcommand {\half} {\tiny{\frac{1}{2}}} 
\newcommand{\inverse}[1]{{\textstyle\frac{1}{#1}}}

\newcommand{\Sx}{\mathbb{S}}

\newcommand{\Sb}{\mathbf{S}}

\begin{document}

\title{Probability-free models in option pricing: statistically indistinguishable dynamics and historical vs implied volatility }
%
\author{
Damiano Brigo\thanks{Department of Mathematics, Imperial College, London}
\\ \hspace{1cm} \\
{\large Paper presented at the conference} \\ {\large \emph{ Options: 45 Years after the publication of the Black-Scholes-Merton Model}}\\ {\large Jerusalem, 4-5 December 2018.  }}
\date{First version: April 3, 2019. This version: August 9, 2021}

\maketitle

\begin{abstract}
We investigate whether it is possible to formulate option pricing and hedging models without using probability. We present a model that is consistent with two notions of volatility: a historical volatility consistent with statistical analysis, and an implied volatility consistent with options priced with the model. The latter will be also the quadratic variation of the model, a pathwise property. This first result, originally presented in Brigo and Mercurio (1998, 2000) \cite{bm1998,bm2000}, is then connected with the recent work of Armstrong et al (2018, 2021) \cite{abbc,abbc21}, where using rough paths theory it is shown that implied volatility is associated with a purely pathwise lift of the stock dynamics involving no probability and no semimartingale theory in particular, leading to option models without probability.  Finally, an intermediate result by Bender et al. (2008) \cite{bender} is recalled. Using semimartingale theory, Bender et al. showed that one could obtain option prices  based only on the semimartingale quadratic variation of the model, a pathwise property, and highlighted the difference between historical and implied volatility. All three works confirm the idea that while historical volatility is a statistical quantity, implied volatility is a pathwise one. This leads to a 20 years mini-anniversary of pathwise pricing through 1998, 2008 and 2018, which is rather fitting for a talk presented at the conference for the 45 years of the Black, Scholes and Merton option pricing paradigm. 
\end{abstract}

\bigskip

\noindent\textbf{AMS Classification Codes}: 62H20, 91B70 \newline

\bigskip


\pagestyle{myheadings} \markboth{}{{\footnotesize  Damiano Brigo, Probability-free models in option pricing \& historical vs implied volatility}}


\section{The initial question: statistical estimation and valuation}

In this work we focus initially on the following question. Take two models $S$ and $Y$ of stock price dynamics under the objective (or statistical / historical / physical) probability measure ${\mathbb{P}}$. Fix a discrete time trading grid, starting from time $0$ and up to a final time $T>0$, with time step $\Delta$. $\Delta$ can be as small as needed but it has to be fixed in advance, at time $0$. 
We then consider pricing options on the stock, according to either model $S$ or $Y$, via the continuous time theory of Black, Scholes and Merton (BSM) \cite{blackscholes},\cite{merton} and subsequent extensions, especially by Harrison et al. \cite{harrison&kreps}, \cite{harrison&pliska}. 
Our question is the following. Can we find situations where $S$ and $Y$ are statistically very close or even indistinguishable (under $\mathbb{P}$), having very close laws in the $\Delta$ grid, but where they imply very different option prices? Prices will be computed as expectations of discounted cash flows under the pricing (or risk-neutral/martingale) measure $\mathbb{Q}$. 
If we do find such situations, can we do this in a constructive way, rather than just proving they exist?


\section{Indistinguishable models leading to different option prices}

To answer this question, we begin with two different models $S$ and $Y$. We take $S$ as the Black Scholes and Merton model, and we construct a second model $Y$ whose marginal laws are the same as $S$. 

\subsection{Matching margins}

Start from the Black-Scholes-Merton model for the stock price $S$ given by 
\[ dS_t= \mu S_t dt + \sigb S_t dW_t\] (abbreviated BSM$(\mu,\sigb)$) with initial condition $S_0=s_0$ under the objective measure $\mathbb{P}$. Here $W$ is a Brownian motion under $\mathbb{P}$, while $s_0>0$, $\mu$ and $\sigb>0$ are constants.  

We then look for a process $Y$, \[dY=u(Y,\ldots)dt + \sigma_t(Y_t) dW_t \] with local volatility $\sigma$ and with the same margins as $S$, namely $p_{S_t} = p_{Y_t}$ for all $t \in [0,T]$. Here for a random variable $X$ we denote by $p_X$ its probability density function.  To find $Y$, invert the Fokker Planck (FP) equation for $Y$ and find the drift $u$ such that the FP equation for the density of $Y$ has solution equal to $p_{S_t}$, namely the lognormal density of the original $S$. 

This was done in Brigo and Mercurio (1998, 2000) \cite{bm1998,bm2000} using previous results on diffusions with laws on exponential families (Brigo (1997) \cite{brigogyor} and Brigo (2000) \cite{brigo00}). 
 
We obtain the following model for $Y$. To avoid singularities of our model coefficient $u$ near $t=0$, we start with a regularization in a small interval $[0,\epsilon)$ that has the same dynamics as $S$, and then we move to the different dynamics from time $[\epsilon,T]$. We obtain


\begin{eqnarray} \label{sol:bes1}
d \bar{Y}_t &=& \mu \bar{Y}_t \ dt + \sigb \bar{Y}_t dW_t, \ 0 \le t <\epsilon, \  \bar{Y}_0 = s_0, \nonumber  
\\
Y_t &=& \bar{Y}_t  \ \mbox{for} \ t \in [0,\epsilon),  \nonumber \\
d Y_t &=& u^\sigma_t(Y_t,s_0,0) dt + \sigma_t(Y_t) dW_t,
\ \ Y_\epsilon = \bar{Y}_{\epsilon^-}, 
\ \  \epsilon \le t \le T,  \\ \nonumber  \\
\nonumber
 u^\sigma_t(x,y,\alpha) &:=& \half \frac{\partial (\sigma_t^2)}{\partial x}(x) +
\half \frac{(\sigma_t(x))^2}{x} \left[ \frac{\mu}{\sigb^2} - \frac{3}{2} -   \frac{1}{ \sigb^2 (t-\alpha)}
\ln\frac{x}{y}\right]\\  &+& \frac{x}{2(t-\alpha)}
\left[\ln\frac{x}{y} - \frac{\frac{\mu}{\sigb^2} - \frac{1}{2}}{2 - \frac{1}{2 \sigb^2 (t-\alpha)}}\right]. \nonumber
\end{eqnarray}
We have introduced $\bar{Y}$ to avoid singularities in the drift coefficient of the SDE \eqref{sol:bes1} near $t=0$. 

The process $Y$, if the related SDE has a solution that is regular enough and admits  densities, has the same marginal distribution as BSM$(\mu, \sigb)$: $p_{S_t} = p_{Y_t}$ for all $t$. We will show a fundamental example where everything works fine in Section \ref{sec:fundam} below. 


\subsection{Matching the whole law on a $\Delta$ grid}

For our purposes of statistical indistinguishability, the above $Y$ is not enough. 
A further fundamental
property of the BSM$(\mu,\sigb)$ model is that its log-returns satisfy
\begin{eqnarray*} 
 \ln \frac{S_{t+\delta}}{S_t}
\sim {\cal N}\left((\mu - \half \sigb^2)\delta ,\ \sigb^2 \delta\right),
\ \ \delta > 0, \ \ t\in [0,T-\delta].
\end{eqnarray*}
Alternative models such as our $Y$ above do not share this
property because identity of the marginal laws alone
does not suffice to ensure it. We need equality of
second order laws or of transition densities.

To tackle this issue, we restrict the set of dates for
which the log-return property
 must hold true. Modify the
definition of $Y$ so that, given 
${\cal T}^\Delta:=\{0,\Delta, 2\Delta, \ldots, N\Delta\}$,
$\Delta = T/N$, $\Delta > \epsilon$, we have
\begin{eqnarray} \label{yret}
 \ln \frac{Y_{i \Delta}}{Y_{j \Delta}}
\sim {\cal N}((\mu - \half \sigb^2)(i-j) \Delta ,\ \sigb^2(i-j) \Delta),
\ \ i > j.
\end{eqnarray}
Limiting such key property to a finite set of times is not so
dramatic. Indeed, only discrete time samples are observed in practice,
so that once the time instants are fixed, our  process $Y$ can not be
distinguished statistically from the Black-Scholes process $S$. 

The new definition of $Y$ we propose now, to match log returns distributions in grids, is still based on our earlier $Y$. 
However, we use the earlier $Y$ process locally in each time interval $[(i-1)\Delta, \ i\Delta)$. In such interval we define iteratively the drift $u^\sigma$ as in the
earlier $Y$ but we translate back the time--dependence of a time amount $(i-1)\Delta$, thus locally restoring the dynamics of the original result for margins starting from time $0$, and we replace $Y_0$ with the final value of $Y$ in the previous interval. We obtain, in each interval $[i \Delta, (i+1) \Delta)$:
\begin{eqnarray}
\label{sol:bes2}
&& d \bar{Y}_t = \mu \bar{Y}_t dt + \sigb \bar{Y}_t dW_t,  \ \   t \in [i\Delta,i\Delta+ \epsilon), \ \ \bar{Y}_{i \Delta} = Y_{i \Delta^-} \nonumber \\
&& Y_t = \bar{Y}_t \ \mbox{for} \ t \in [i\Delta,i\Delta+ \epsilon), \nonumber \\
&& d Y_t = u^\sigma_t(Y_t,Y_{\alpha(t)},\alpha(t)) dt +  \sigma_t(Y_t) dW_t, \ \
t \in [i\Delta + \epsilon, (i+1)\Delta), \ \ Y_{i \Delta + \epsilon} = \bar{Y}_{i \Delta + \epsilon^-}   
\end{eqnarray}
where $\bar{Y}_0 = Y_0 = s_0$,  $u^\sigma_t(x,y,\alpha)$ was defined in the earlier $Y$ and  $\alpha(t) = i \Delta \ \mbox{for}  \  t \in [i\Delta, \ (i+1)\Delta)$.

It is clear that the transition densities
of $S$ and $Y$ satisfy
$p_{Y_{(i+1) \Delta}|Y_{i \Delta}}(x;y) = p_{S_{(i+1) \Delta}|S_{i\Delta}}(x;y)$ by construction.

Note also that the new process $Y$ is not a Markov process
in $[0, T]$. However, it is a Markov process in all time instants of ${\cal T}^\Delta$
($\Delta${\em --Markovianity}).

Finally, note that now the two models $S$ (BSM$(\mu,\sigb)$) and $Y$ are {\emph{statistically indistinguishable}} in ${\cal T}^\Delta$ since there they share the same finite dimensional distributions. Any statistician who tried to estimate the two models from data could not find a way to distinguish them. Before turning to option prices implied by the two indistinguishable models, we need to show that we have not produced an empty theory so far. In other terms, we need to give concrete examples of $\sigma$ for which our framework works rigorously. Such a fundamental case is addressed in the next section. 


\subsection{A fundamental case: $\sigma_t(y) = \nu y$}\label{sec:fundam}
We take now $\sigma(Y) = \nu Y$, so that also the volatility of $Y$ is of BSM type, but with constant  $\nu$ instead of $\sigb$. 
In this case the equation for $u$ specializes to
\[ u^{\sbar}_t(y,y_\alpha,\alpha) =
y \left[ \inverse{4}(\sbar^2 - \sigb^2) +
\frac{\mu}{2}(\frac{\sbar^2}{\sigb^2} + 1) \right]
+ \frac{y}{2(t-\alpha)} (1-\frac{\sbar^2}{\sigb^2} ) \ln\frac{y}{y_\alpha},\]
and in this fundamental case one can show that the SDE for $Y$ has a unique strong solution \cite{bm1998}, \cite{bm2000}. 

Moreover, the change of measure that replaces the drift $u$ with $r Y$ is well defined and regular, so that it is possible to change probability measure from ${\mathbb P}$ to the equivalent pricing measure  ${\mathbb{Q}}$ for the model $Y$. 

This is precisely what we are interested in, since changing measure leads to some quite interesting developments. Before turning to the change of measure, a final remark concerning the regularization in $[0,\epsilon)$ is in order.

\subsection{A technical link with the rough volatility literature}

As $Y$ has the same margins 
as $S$, it has to be positive like $S$, so that $Y_t>0$. Then take $Z_t^\epsilon := \ln Y_t$: 
\begin{eqnarray} \label{intediff}
Z^\epsilon_t &=& Z^\epsilon_{j\Delta} + (\mu -\half\sigb^2)(t-j\Delta)  \\ \nonumber
& &
+\left\{
\begin{array}{ll}
\sigb(W_t-W_{j\Delta}) \ \ \ \  \ \mbox{for} \ t\in [j\Delta,j\Delta+\epsilon), &  \\ \nonumber
\left(\frac{t-j\Delta}{\epsilon}\right)^{\beta/2}
\left[\sigb(W_{j\Delta+\epsilon}-W_{j\Delta})+ \sbar\int_{j\Delta+\epsilon}^t
\left(\frac{u-j\Delta}{\epsilon}\right)^{-\beta/2} dW_u\right] \ \mbox{for} \ t \in [j \Delta + \epsilon, (j+1)\epsilon).
& 
\end{array}
\right.
\end{eqnarray}
Here $\beta = 1-\frac{\sbar^2}{\sigb^2}$.
In \cite{bm1998} we show that we can take $\epsilon \rightarrow 0$ in the regularization, obtaining a limit $Z$ 
\begin{eqnarray*}
Z_t = Z_{j\Delta} + (\mu -\frac{\sigb^2}{2})(t-j\Delta)  +
   \sbar\int_{j\Delta}^t
 \left[\frac{t-j\Delta}{u-j\Delta}\right]^{\frac{\beta}{2}} dW_u,
\ \ \  t\in [j\Delta ,(j+1)\Delta).
\end{eqnarray*}
This process is well defined since the integral in the right-hand side exists finite almost surely 
even though its integrand diverges when $u \rightarrow j \Delta^+$.
%
%
%
The above equation can be better compared to the Black and Scholes
process when written in differential form:
\begin{eqnarray*}
d Z_t = (\mu -\half\sigb^2)\ dt +
\frac{\beta}{2}(t-j\Delta)^{\beta/2-1}  \left(\int_{j\Delta}^t (u-j\Delta)^{-\beta/2} dW_u\right)\sbar\ dt
  + \sbar \ dW_t  \ \  \  t\in [j\Delta ,(j+1)\Delta).
\end{eqnarray*}
The central term in the right hand side is  needed to have returns equal to the Black and Scholes process even after changing the volatility coefficient  
from $\sigb$ to $\sbar$. More precisely, the central term is the correction needed so that  the exponential of $Z$ will simultaneously have returns equal to those of BSM$(\mu,\sigb)$ and volatility coefficient equal to $\sbar$. Note that this term goes to zero for $\sigb = \sbar$.
It is this term that makes our process non-Markov outside the trading time grid.

Finally, we point out that for all $\epsilon >0$ there exists a probability measure $\mathbb{Q}^\epsilon$ equivalent to $\mathbb{P}$ and such that $Z^\epsilon$ is a Brownian motion under $\mathbb{Q}^\epsilon$. As noted in \cite{bellanipc}, informally taking the limit for $\epsilon \downarrow 0$ in such change of measure yields a ``market price of risk'' for the discrepancy between historical and implied volatilities. Such market price of risk has two distinguished properties that have been important in the rough volatility literature, see for example \cite{bayer}: small time behaviour proportional to $t^{-\gamma}$ for some $0<\gamma<1$, and rougher trajectories than semimartingale ones.

%


\subsection{Arbitrarily different option prices under indistinguishable models}
We now consider option pricing based on the models $S$ and $Y$. We will assume that interest rates $r$ are deterministic and constant in time. 

Recall our two indistinguishable models under the measure $\mathbb{P}$, and see what happens when we change measure to $\mathbb{Q}$:
\begin{eqnarray*}  
 d S_t = \mu S_t dt + \sigb S_t dW^{\mathbb{P}}_t  &\Delta-\mbox{indistinguishable from}& d Y_t = u^{\sbar}_t dt + \sbar \ Y_t \ dW^{\mathbb{P}}_t, \\  \\
 &\mbox{but changing measure to} \ \mathbb{Q}& \\ \\
 d S_t = r S_t dt + \sigb S_t dW^{\mathbb{Q}}_t\   &\mbox{very\ different\  from}& d Y_t^\sbar = r Y^\sbar_t dt + \sbar \ Y_t \ dW^{\mathbb{Q}}_t\ 
 \end{eqnarray*}
No arbitrage conditions are reflected in the fact that changing measure to $\mathbb{Q}$ enforces the drift $r$.  If we now price a call option with the $\mathbb{Q}$ expectation of the discounted payoff  we have
\[ \mathbb{E}^{\mathbb{Q}}[e^{-rT} (S_T -K)^+]=\mbox{BlackScholesFormula}(\sigb), \ \ 
 \mathbb{E}^{\mathbb{Q}}[e^{-rT} (Y_T -K)^+]=\mbox{BlackScholesFormula}(\sbar) \]
where the remaining inputs of the Black and Scholes formula, namely $s_0, r, T$ are the same for both cases. 
Since the indistinguishability holds for every $\nu$, we can take $\nu\downarrow 0$ and $\nu \uparrow +\infty$. This way we find the following.

\bigskip
 
{\emph{Statistically indistinguishable stock price models in the $\Delta$ time grid imply option prices so different to span the whole no arbitrage interval}} $[(S_0 -K e^{-rT})^+, S_0]$.   

\bigskip

Perhaps surprisingly, they span a range that is not related to $\Delta$. Since models are equivalent in a $\Delta$ grid, one might have expected that by tightening $\Delta$ one might have had  option prices in a narrower range. This is not the case.  


Our result shows that conjugating discrete and continuous time modeling (e.g. statistics and option pricing) has to be done carefully and is subject to important assumptions.


\section{Reconciling historical and implied volatility with a single dynamics}


What if what we have seen is a way to account consistently for {\emph{historical}} and {\emph{implied}} volatility? Indeed, it is known that option prices trade independently of the
underlying stock price, and we have been able to
construct a stock price process $Y^{\nu}$ whose marginal
distribution and transition density depend on the volatility coefficient $\sigb$ (historical volatility),
whereas the corresponding option price only depends on the volatility coefficient $\nu$ (implied volatility). As a consequence, we can  provide a
consistent theoretical framework justifying the differences between
historical and implied volatility that are commonly observed in real markets.



\section{Possible explanation of arbitrary option prices?}

Since probability and statistics have proven to be deceptive when working in discrete time  under ${\mathbb{P}}$, we try now to remove probability and statistics from valuation. We will achieve this by using ideas from rough paths theory. 
Rough paths theory has been initiated and developed by many researchers over the years, here we recall briefly first  F\"ollmer \cite{foellmer}, who introduced a form of It\^o calculus without probability. A more general theory was introduced and developed in Lyons \cite{lyons98}, Davie \cite{davie}, Gubinelli \cite{gubinelli} and Friz among others, see for example \cite{frizhairer} and references therein. 
Applying ideas from rough paths theory, and from \cite{davie,gubinelli,frizhairer} in particular, in Armstrong et al. (2018, 2021) \cite{abbc,abbc21} we manage to re-interpret the Black Scholes formula and option pricing in a purely pathwise sense. 

%
%
After writing \cite{abbc} we found out that there had been earlier attempts to build a pathwise theory of option pricing. Bender et al. (2008) \cite{bender} had formulated option pricing, in the framework of semimartingale theory, relying only on the quadratic variation of price trajectories, which is a {\emph{pathwise property}}. In our work \cite{abbc, abbc21} the analysis is brought one step further by abandoning the semimartingale setting. Using Davie's rough differential equations and {\emph{rough brackets}} (see \cite{frizhairer}) we  abandon probability theory altogether. 
This entails the definition of Gubinelli derivatives, which in the classical Black and Scholes’ framework are Gamma sensitivities of the options. Although ignored in the most classical formulas for portfolio dynamics, they play an active role in the convergence of the integrals describing portfolio processes, when such convergence is analysed pathwise as in \cite{abbc, abbc21}. 
We should also say that besides \cite{bender}, there had been earlier attempts at a pathwise theory of option pricing. Such attempts are based on the above-mentioned non-probabilistic approach to It\^o calculus by F\"ollmer (1981) \cite{foellmer} and are given in Bick \& Willinger (1994) \cite{bick}, see also the more recent work of Schied \& Voloshchenko (2016) \cite{schied}. These approaches do not emphasize the role of Gamma as a Gubinelli derivative nor do they hold under the generality of rough path theory. 
Indeed, F\"ollmer calculus has the caveat that the integrals depend upon the discrete-time approximating sequence, which precludes obtaining important robustness results we illustrate in \cite{abbc,abbc21}. To bypass this problem, in \cite{abbc,abbc21} we use an augmented version of delta-hedging where one also invests in volatility swaps to hedge the second-order part of the pricing signal. This produces a robust trading strategy but at the cost of introducing assumptions on the price of volatility swaps to guarantee our strategy is self-financing.
Furthermore, the above-mentioned pathwise formulations require paths that can be as rough as  semimartingales but not rougher, namely paths of of finite $p$-variation for all $p>2$. Using Rough Path Theory, we can accommodate paths of finite $p$-variation with $2<p<3$, thus showing that delta hedging can be extended to a broader class of price signals.

%
%
We now go back to the statistically indistinguishable models $S$ and $Y$ above, implying arbitrarily different option prices. We connect them to our discussion of pathwise properties. Since the results from \cite{bender} and \cite{abbc,abbc21} suggest that option pricing ultimately depends only on pathwise features of price trajectories, it is with the lenses of pathwise analysis that we can distinguish the $\mathbb{P}$ dynamics of $S$ and of $Y$, by looking at their quadratic variation. While probability and statistics could not distinguish between $S$ and $Y$, the difference is instead captured by this aspect of price trajectories. 

Paraphrasing:  
\begin{itemize}
\item Probability and Statistics do not allow one to distinguish between $S$ and $Y$ in ${\cal T}^\Delta$.
\item Prices of options written on $S$ and $Y$ are different, so that option prices allow one to  distinguish between $S$ and $Y$.
\item One then conjectures that option prices cannot be properties of $S$ and $Y$ based on probability or statistics. With prices being traditionally associated with expectations of discounted cash flows, this seems initially counterintuitive.  
\item \cite{abbc,abbc21} show that, in general, option prices can be derived based only on purely pathwise properties, without using probability theory anywhere and in a robust way, with paths that than be rougher than semimartingales.  \cite{bender,bick,schied} had obtained similar results earlier but within semimartingale theory or within paths with the same regularity as semimartingales. 
\item In particular, in models $S$ and $Y$ option prices are completely specified by probability-free pathwise properties of $S$ and $Y$, confirming the above conjecture.   
\end{itemize}
 
We now summarize how in \cite{abbc,abbc21} we were able to give a probability-free, and more specifically semimartingale-free derivation of option pricing based on purely pathwise properties.



\subsection{Rough paths and option pricing}

We denote in general $X_{s,t} = X_t - X_s$. Recall the BSM$(\mu,\sigb)$ model 
\begin{eqnarray*}
dS_t=S_t[\mu dt+ \sigb dW^{\mathbb P}_t], \ \ \ dB_t = r B_t  dt, \ \ \ 0\leq t \leq T
\end{eqnarray*}
where we also included the risk-free bank account numeraire $B$. In the classic theory of stochastic differential equations (SDEs), the above equation is a short form for an integral equation
\[ S_t - S_0 = \int_0^t \mu S_u du + \int_0^t \sigb S_u dW_u . \]
The last integral is an Ito stochastic integral. If one tries to re-formulate the above equation without probability, one will not be able to use stochastic integrals any more, and as a result one will need to define the integral $\int_0^t \sigb S_u dW_u$ pathwise. To do this, one will need to add information on the price trajectory in the form of a lift. One needs to provide the input 
\[ \Sx_{s,t} = \int_{s}^t S_{s,u} dS_u .\]
This is really an input and is not defined a priori based only on properties of $S$: if the signal $S$ has finite $p$-variation for $2<p<3$, as in case of paths in the Black Scholes model, it is too rough to define the above intergral as a Stiltjes or Young integral. One needs therefore to add $\Sx$ as an input.  
To understand why this is important, we now explain how introducing $\Sx$ helps in defining integrals of the type $\int F(S_r) dS_r$. 
Consider $\int F(S_r) dS_r$ and try to write it as a Young integral.
Take Taylor expansion $F(S_r) \approx F(S_u) + D F(S_u) S_{u,r}$.
The Young integral can be seen as approximating $F(S_r)$, in each $[u,t]\in \pi$ with the zero-th order term $F(S_u)$, where $\pi$ is the partition for the discrete sums approximating the integral, and $|\pi|$ is the mesh size. 
Hence 
\[\int_0^T F(S_r) dS_r = \lim_{|\pi|\rightarrow 0} \sum_{[u,t] \in \pi} \int_u^t F(S_u) dS_r  =  
\lim_{|\pi|\rightarrow 0} \sum_{[u,t] \in \pi} F(S_u) S_{u,t} .\]

The limit is on {\emph{all partitions}} whose mesh size tends to zero. 
If we cannot use a Young integral because $S$ is too rough, we can try a first order expansion for $F(S)$ rather than a zero-th order one.
\begin{eqnarray}\label{eq:roughint} \int_0^T F(S_r) d\Sb_r   &=&  
\lim_{|\pi|\rightarrow 0} \sum_{[u,t] \in \pi}  \int_u^t \left(F(S_u) + D F(S_u) S_{u,r}\right) dS_r = \nonumber \\ &=&
\lim_{|\pi|\rightarrow 0} \sum_{[u,t] \in \pi} (F(S_u) S_{u,t} + D F(S_u) \boxed{\Sx_{u,t}} ) . 
\end{eqnarray}
This intuition can be made rigorous with the relevant norms and metrics, see for example \cite{frizhairer}. 

What attracted initially the interest of the author, prompting the initial work that later lead to the collaboration culminating in \cite{abbc,abbc21}, is the fact that in a delta-hedging context, one can write the replication condition for a call option in a Black-Scholes type model as
\[ \int_0^T \Delta(r,S_r) d S_r +  \int_0^T \eta(r,S_r) d B_r  = (S_T-K)^+ - V_0 \] 
where $V_0$ is the option price at time $0$ and $\Delta$ and $\eta$ are the amounts of stock and cash in the self-financing replicating strategy. In particular, $\Delta$ is the sensitivity of the option price with respect to its underlying $S$. The above integral $\int_0^T \Delta(r,S_r) d S_r$ is an Ito integral. If one tries to re-write the replication condition using a purely pathwise integral, using rough integration as in \cite{frizhairer}, it is possible to do so by recalling that, according to \eqref{eq:roughint}, 
\[ \int_0^T \Delta(r,S_r) d \Sb_r  = \lim_{|\pi|\rightarrow 0} \sum_{[u,t] \in \pi} (\Delta(u,S_u) S_{u,t} + D_S \Delta(u,S_u) \Sx_{u,t} ) . \]
In an option pricing setting, the term $D_S \Delta(u,S_u) =: \Gamma(u,S_u)$ turns out to be the second derivative of the option price with respect to its underlying term, the so called gamma of the option. Hence
\[\lim_{|\pi|\rightarrow 0} \sum_{[u,t] \in \pi} (\Delta(u,S_u) S_{u,t} + \Gamma(u,S_u) \Sx_{u,t} ) +  \int_0^T \eta(r,S_r) d B_r  = (S_T-K)^+ - V_0 .\] 
This immediately prompted the author to notice that gamma played an explicit role in the replication condition when the condition is expressed using purely pathwise integrals rather than Ito integrals. Traders have been always using gamma as a correction term, but the above limit shows a fundamental role for gamma already at the level of the replication condition. On the contrary, the Ito integral  $\int_0^T \Delta(r,S_r) d S_r$ can be written as a limit where gamma does not appear. 
The author further proposed to interpet the lift $\Sx$ as related to a non-standard covariance swap. This is discussed further in \cite{abbc,abbc21} and will be expanded in future work. 

Let us now go back to BSM and \cite{abbc,abbc21}. Before proceeding further, we need to explain that in \cite{abbc,abbc21} we do not use really full rough path theory. We presented above a sketch of how one makes sense of the BSM SDE for $S$ in a purely pathwise way, but in effect we never use the full SDE. Indeed, we noted in the discussion above that only the quadratic variation of the SDE solution matters in determining the option price. Similarly, in the no-semimartingales pathwise case, we will not really need the pathwise analogous of the BSM SDE, but just a no-semimartingale analogous of the quadratic variation. More precisely, take $S_t$ as a path of finite $p$ variation with $2<p<3$ (Brownian motion has finite $q$ variation for all $q > 2$, so $S$ is potentially rougher than BSM). 
Consider the lifted $\Sb_t:= (S_t,\Sx_t)$, where $\Sx$ is our input for $\int S \ dS$. 
From a technical point of view in \cite{abbc,abbc21} we work with {\emph{reduced}} rough paths, obtained from the pair $(S,\Sx)$ by considering only the symmetric part of $\Sx$ (this distinction is essential in the multi-dimensional case, although here we are discussing the one-dimensional setting). This is equivalently described by the rough bracket defined in 
\[ [ \Sb ]_{u,t} =S_{u,t} S_{u,t}-  2\ \Sx_{u,t} .\]
We refer again to \cite{frizhairer} for the details, and point out the early work of F\"ollmer \cite{foellmer} in this regard.
%



%
%
%

If $[\Sb]_{u,t}$ is regular enough to define a measure of $[u,t]$ with density $a(S_t)$ with $a(x)$ also regular, then the classic partial differential equation for the option price is defined entirely in terms of the  {\emph{purely pathwise}} bracket $[\Sb]$, involving no probability theory and no semimartingale theory in particular. It follows that the option price itself will not depend on the probabilistic setting but only on path properties. 

The purely pathwise property $[\Sb]_{u,t}$ takes the place of implied volatility in determining the option price as a path property rather than a statistical property. The latter would be associated with historical volatility as a standard deviation (statistics).


\section{Conclusions: 20 years of pathwise pricing}


The first result we reported in this paper is the result of Brigo and Mercurio (1998) \cite{bm1998}, where the process $Y$ had simultaneously historical volatility $\sigb$ as a statistical property and implied volatility $\nu$ as a pathwise property (quadratic variation). This is consistent with the later result of  Bender, Sottinen and Valkeila (2008) \cite{bender} who note:

\bigskip
 

{\emph{``[...] the covariance structure of the stock returns is not relevant for option pricing, but the quadratic variation is.  So, one should not be surprised if the historical and implied volatilities do not agree: the former is an estimate of the variance and the latter is an estimate of the [semimartingale] quadratic variation''.}}

\bigskip

This has been further generalized in Armstrong et al. (2018, 2021) \cite{abbc,abbc21} where historical volatility is a statistics of the variance too, while implied volatility  is associated with a pathwise lift involving no semimartingale theory and no probability.

\bigskip

It is perhaps fitting that this 20 years anniversary of pathwise pricing, running through 1998, 2008 and 2018, is occurring at the conference on the 45th anniversary of the Black, Scholes and Merton option pricing theory, even though we should not forget earlier contributions like \cite{bick} and other recent contributions such as \cite{schied}.

%
%
%
%
%
%

\section*{Acknowledgments}
The author is grateful to Claudio Bellani for checking the draft of the paper and for many helpful suggestions and to Mikko Pakkanen for referring him to the work of Bender, Sottinen and Valkeila.  The author is further grateful to the organizers and participants of the conference ``Options: 45 Years after the publication of the Black-Scholes-Merton Model'', held in Jerusalem on 4--5 December 2018, for their comments and suggestions. 
 
%

 




\begin{thebibliography}{99}

\bibitem{abbc} Armstrong, J., Bellani, C., Brigo, D., and Cass, T. (2018). 
Gamma-controlled pathwise hedging in generalised Black-Scholes models.
{\tt{https://arxiv.org/abs/1808.09378v1}}\\  Updated in April 2019 with the title 
``Option  pricing models without probability'', {\tt{https://arxiv.org/abs/1808.09378v2}}


\bibitem{abbc21} Armstrong, J., Bellani, C., Brigo, D., and Cass, T. (2021). 
Option pricing models without probability: A rough paths approach. Mathematical Finance, 2021, 1-- 28. {\tt{https://doi.org/10.1111/mafi.12308}}


\bibitem{bayer}
Bayer, C., Friz, P. and Gatheral, J. (2016). Pricing under rough volatility. Quantitative Finance, 16(6):1--18, 2016.

\bibitem{bellanipc} Bellani, C. (2018). Connection between the result in Brigo Mercurio (2000)   and rough volatility. Internal note, Imperial College London. 

\bibitem{bender} Bender, C., Sottinen, T., and Valkeila, E. (2008).  Pricing by hedging and no-arbitrage beyond semimartingales. Finance and Stochastics, Vol. 12(4), pp 441--468

\bibitem{bick} 
Bick, A., and Willinger, W. (1994). Dynamic spanning without probabilities. Stochastic Process. Appl., 50(2), 349–374.

\bibitem{blackscholes} Black, F. and Scholes, M. (1973). 
The Pricing of Options and Corporate Liabilities. The Journal of Political Economy, Vol. 81, No. 3, pp. 637--654

\bibitem{brigogyor}
Brigo, D. (1997). On nonlinear {SDEs} whose densities evolve in a finite--dimensional
  family. In: Stochastic Differential and Difference Equations, Progress in
  Systems and Control Theory 23: 11--19, Birkh{\"{a}}user, Boston.

\bibitem{brigo00}    Brigo, D (2000). On SDEs with marginal laws evolving in finite-dimensional exponential families, Statistics and Probability Letters, 49: 127 -- 134


\bibitem{bm1998} Brigo, D. and Mercurio, F. (1998). 
Discrete time vs continuous time stock price dynamics and implications for option pricing. arXiv.org and SSRN.com

\bibitem{bm2000} Brigo, D. and Mercurio, F. (2000). 
Option pricing impact of alternative continuous time dynamics for discretely observed stock prices. Finance \& Stochastics, 4, pp. 147-159

\bibitem{davie}
Davie, A. M. (2007). Differential equations driven by rough paths: an approach via discrete approximation. Appl. Math. Res. Express AMRX 2, 40.  


\bibitem{foellmer} F\"ollmer, H. (1981). Calcul d'Ito sans probabilit\'es. S\'eminaire de probabilit\'es de Strasbourg 15 (1981): 143--150. 

\bibitem{frizhairer} Friz, P., and Hairer, M. (2015). A course on rough paths. Springer Verlag, Heidelberg.


\bibitem{gubinelli} Gubinelli, M. (2004). Controlling rough paths. J. Funct. Anal. 216 (2004) 86–140.

\bibitem{harrison&kreps} Harrison, J.M, and Kreps, D.M. (1979). Martingales and
arbitrage in multiperiod securities markets, Journal of Economic
Theory, 20(3), pp. 381--408

\bibitem{harrison&pliska} Harrison, J.M., and Pliska, S.R.  (1981).  Martingales and
Stochastic Integrals in the Theory of Continuous Trading,
Stochastic Processes and their Applications, 11(3),pp. 215--260.


\bibitem{lyons98}
Lyons, T.J. (1998). 
\newblock Differential equations driven by rough signals.
\newblock { Rev. Mat. Iberoamericana}, 14(2):215--310.


\bibitem{merton} Merton, R. C. (1973). Theory of Rational Option Pricing. Bell Journal of Economics and Management Science. The RAND Corporation. 4 (1): 141--183

\bibitem{schied} Schied, A., and Voloshchenko, I. (2016). Pathwise no-arbitrage in a class of delta hedging strategies. Probability, Uncertainty and Quantitative Risk, 1(1), 1--25.




\end{thebibliography}
 \end{document}